# Distribution Locational Marginal Pricing Under Uncertainty Considering Coordination of Distribution and Wholesale Markets

Zongzheng Zhao, Yixin Liu, Li Guo, *Member, IEEE*, Linquan Bai, *Senior Member, IEEE*, and Chengshan Wang, *Senior Member, IEEE*

***Abstract*—An effective distribution electricity market (DEM) is required to manage the rapidly growing small-scale distributed energy resources (DERs) in distribution systems (DSs). This paper proposes a day-ahead DEM clearing and pricing mechanism to account for the uncertainty of DERs and the coordination with the wholesale electricity market (WEM) through a bi-level model. The upper-level model clears the WEM in the transmission system (TS) and forms the locational marginal price (LMP) and uncertainty LMP (ULMP) for energy and uncertainty/reserve, respectively. In the lower level, a robust scheduling model considering WEM-DEM coordination and uncertainties is proposed to clear the DEM. Accordingly, the distribution LMPs (DLMPs) for active power, reactive power and uncertainty/reserve are derived to reward the energy/reserve provision and charge uncertain resources in the DEM, which provide effective price signals for managing not only the voltage and congestion, but also the uncertainty in DSs. A heterogeneous decomposition (HGD) algorithm is utilized to solve the bi-level model in a decentralized manner with limited information interaction between TS and DSs, which guarantees the solution efficiency and information privacy. The effectiveness of the proposed method is verified via numerous case studies.**

***Index Terms*—Coordination, distribution electricity market (DEM), distribution locational marginal pricing (DLMP), robust optimization, uncertainty, wholesale electricity market (WEM).**

## NOMENCLATURE

### *Indices, Sets, and Parameters*

*Transmission Side:*

| | |
|---|---|
| $\mathcal{G}_\mathcal{T}$,$\mathcal{W}_\mathcal{T}$, $\mathcal{D}_\mathcal{T}$,$\mathcal{DS}_\mathcal{T}$ | Set of thermal generators (TGs), wind farms (WFs), load serving entities (LSEs), and DSs |
| $\mathcal{N}_\mathcal{T}$ | Set of buses |
| $P^W_{i,t}$,$P^D_{i,t}$ | Forecast value of WF/LSE $i$ at time $t$ |

*Distribution Side:*

| | |
|---|---|
| $\mathcal{DER}$,$\mathcal{RDG}$ $\mathcal{MT}$,$\mathcal{ESS}$ | Set of DERs, renewable distributed generation (RDG), microturbines (MTs), and energy storage systems (ESSs) |
| $\mathcal{N}$,$\mathcal{L}$ | Set of nodes and distribution lines |
| $pr(i)$,$cr(i)$ | Set of parent and children of node $i$ |
| $P^d_{i,t}$,$Q^d_{i,t}$ | Active/reactive load demand at node $i$ at time $t$ |
| $P^{rdg}_{i,t}$ | Forecast value of RDG $i$ at time $t$ |

### *Variables*

*Transmission Side:*

| | |
|---|---|
| $u_{i,t}$,$v_{i,t}$,$I_{i,t}$ | Startup (1 startup, 0 otherwise), shutdown (1 shutdown, 0 otherwise), and unit commitment (1 online, 0 otherwise) status of TG $i$ at time $t$ |
| $P^G_{i,t}$,$\Delta P^G_{i,t}$ | Energy/reserve provided by TG $i$ at time $t$ |
| $\delta^W_{i,t}$,$\delta^D_{i,t}$ | Forecast deviation of WF/LSE $i$ at time $t$ |

*Distribution Side:*

| | |
|---|---|
| $P^{DS}_{i,t}$,$Q^{DS}_{i,t}$, $R^{DS}_{i,t}$ | Active/reactive power and reserve demands of a DS at bus $i$ in TS and at time $t$ |
| $N^{cb}_{i,t}$ | Number of operating units in CB $i$ at time $t$ |
| $o_{k,t}$ | Position of the tap on OLTC, $o_{k,t} = 1$ if the tap is at the $k$th position, 0 otherwise |
| $\phi_{i,t}$, $\psi_t$ | Status change of CB $i$ or OLTC at time $t$ (1 changed, 0 unchanged), respectively |
| $P^g_{i,t}$,$Q^g_{i,t}$ | Active/reactive power of DER $i$ at time $t$ |
| $\epsilon^{rdg}_{i,t}$ | Forecast deviation of RDG $i$ at time $t$ |

## I. INTRODUCTION

THE rapid growth of distributed energy resources (DERs) accelerates the transition of traditional passive distribution systems (DSs) to active DSs. High-penetration DERs may hinder the normal operation of DSs, causing problems such as overvoltage and network congestion. Establishing a distribution electricity market (DEM) can provide an effective solution for managing large amounts of small-scale DERs in DSs [1].

The DEM establishment faces two major challenges. On one hand, the uncertainty of renewable distributed generation (RDG), e.g., photovoltaics (PVs) and wind turbines (WTs), calls for a cost-effective DEM clearing mechanism that can internalize uncertainties and provide effective price signals to realize uncertainty management and secure operation of DSs. On the other hand, new regulations such as the FERC Order 2222 in the U.S allow DERs with the minimum capacity requirement of 100kW to compete in the wholesale electricity market (WEM) [2]. Thus, a coordination mechanism is required to provide a platform for DERs to participate in not only the DEM, but also the WEM. To this end, this paper proposes to design a DEM clearing and pricing mechanism considering the uncertainty of RDG and WEM-DEM coordination.

Recently, much attention has been paid to the distribution locational marginal pricing (DLMP) mechanism which can implement market clearing and pricing as well as optimize the operation of DERs in the DEM. In [3], the distribution system operator (DSO) calculates DLMPs with the objective of maximizing social welfare to alleviate congestion caused by electric vehicles (EVs). In [4], the DSO publishes DLMPs to aggregators to optimize the scheduling of EVs and heat pumps. In [5], a DLMP method for eliminating congestion is introduced to dispatch prosumers which are responsive to price signals. The above methods adopt DC optimal power flow (DCOPF) in the derivation of DLMPs, thus the reactive power and voltage are neglected. To address this issue, the authors in [6] utilize AC power flow approximation to calculate DLMPs, which can be decomposed into energy, congestion, loss, and voltage components. A day-ahead DEM clearing and pricing model is presented in [1] based on a linearized power flow model considering reactive power and voltage constraints. The studies in [1], [3]-[6] demonstrate the effectiveness of DLMP in DERs management in DSs. However, few studies take into account the uncertainty of RDG in DEM clearing and pricing, which may hinder the application of DLMP in practice and lead to market inefficiencies and economic losses [7].

The uncertainty-aware DLMPs can be obtained by methods such as stochastic optimization, chance-constrained (CC) optimization, and robust optimization (RO). In [8], a stochastic method based on scenario-tree technique is proposed to account for the uncertainty of RDG, and the DLMPs can reflect the influence of RDG on DS operation. In [9], a CC optimal power flow (OPF) is used to internalize the uncertainties into DLMPs. However, the stochastic method is computationally challenging due to the complexity introduced by a large number of scenarios. The CC method requires knowing continuous probability distribution of uncertain variables, which may not be available in practice. In view of these limitations, the RO method is an alternative with significantly reduced computation burden, which minimizes the worst-case cost within an uncertainty set [10]. In [11], DLMPs are derived in day-ahead DEM for congestion management through a RO method to account for the uncertainty of flexible loads. In [12], a RO-based DEM clearing mechanism is presented to calculate DLMPs for pricing active power and uncertainty. However, network loss and the pricing for reactive power are not considered in [11] and [12], which cannot achieve a complete market clearing.

Another important issue in DLMP calculation lies in the coordination between the DEM and WEM. However, most of existing articles on DLMPs, such as [1], [3]-[6], [8], [9], [11], and [12], assume that the boundary locational marginal price (LMP) at the power supply point (PSP) connected to the transmission system (TS) is not impacted by the DSO behaviors. Separate market clearing in TS and DS without coordination may pose some challenges [13]: a) Generation resources may be not fully utilized, which impairs the overall benefits of the TS and DSs. b) Network congestion may be more serious, leading to the increase of LMP. c) Power mismatch may occur at the PSP and affects system operation. Several studies take into account the WEM-DEM coordination to calculate DLMPs [14]-[16]. In [14] and [15], the DLMP calculation based on DCOPF iterates between TS and DSs, until no changes occur in the cleared energy or prices at the PSP. In [16], a bi-level optimization model considering WEM-DEM coordination is presented to clear the DEM and calculate DLMPs through the equilibrium problem with equilibrium constraints (EPEC) approach. However, the uncertainty of RDG as well as its influence on DLMPs are not considered in [14]-[16].

In this paper, a bi-level optimization model is proposed for day-ahead DEM clearing and pricing considering the WEM-DEM coordination and uncertainty of RDG. In the upper level, the transmission system operator (TSO) optimizes energy and reserve schemes based on a robust unit commitment (RUC) model, then forms LMP and uncertainty LMP (ULMP) for energy and uncertainty/reserve, respectively. In the lower level, a robust scheduling model based on linearized power flow is formulated by each DSO to clear the local DEM and calculate DLMPs. The boundary energy/reserve demands of each DS and LMP/ULMP are exchanged at the PSPs between TS and DSs. The bi-level model is solved iteratively by a heterogeneous decomposition algorithm (HGD) in a decentralized manner. The main contributions of this paper are highlighted as follows:

1) A novel DEM clearing and pricing model is proposed to account for the coordination with the WEM. The influence of large amounts of DERs in DSs on WEM clearing and pricing, as well as the influence of LMP/ULMP in TS on DEM clearing and DLMP formation are all considered. The coordination makes it possible for DERs to participate in both DEM and WEM based on the derived DLMPs.

2) A robust scheduling model is formulated to internalize the uncertainty of RDG in DEM clearing and pricing. The reserve schemes of DERs are optimized to deal with uncertainties. All the derived DLMPs for active power, reactive power, and uncertainty (i.e., $\text{DLMP}^{\text{P}}$, $\text{DLMP}^{\text{Q}}$, and $\text{DLMP}^{\text{U}}$) can reflect the uncertainty of RDG, which can provide effective price signals to incentivize the uncertainty management in DSs.

3) A sensitivity-based HGD algorithm is utilized to solve the bi-level model in a decentralized manner. The sensitivity of boundary LMP/ULMP to energy/reserve demands of each DS is obtained from previous input and output by a probing method. The solution convergency and optimality can be guaranteed. In addition, the decentralized method reduces the communication burden and protects the information privacy.

The remainder of this paper is organized as follows: Section II proposes the coordinated WEM-DEM framework. Section III and Section IV present the clearing and pricing models in WEM and DEM, respectively. Section V elaborates the solution methods. Section VI conducts simulation studies and Section VII draws main conclusions for this paper.

## II. Coordinated WEM-DEM Framework

The day-ahead WEM and DEM typically include 24 hours with time resolution of 1 hour. In both markets, it is assumed that all participants bid at marginal cost. The proposed bi-level optimization model is depicted in Fig. 1.

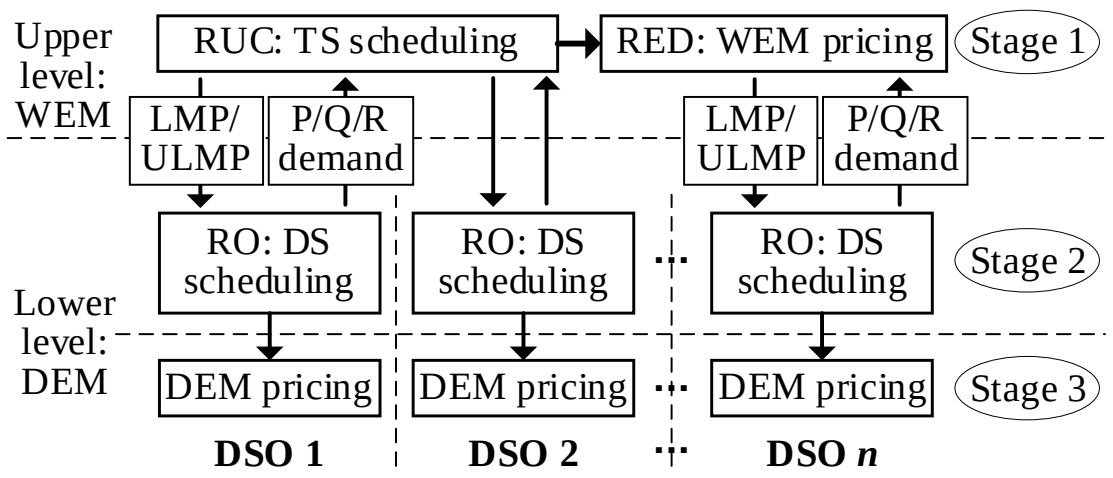

Fig.1. The framework of the bi-level optimization model.

In the upper level (Stage 1), the TSO operates the WEM which includes participants such as wind farms (WFs), thermal generators (TGs), load serving entities (LSEs), and DSs. The WFs and LSEs submit day-ahead forecast value and forecast deviation to the TSO. Meanwhile, each DSO submits day-ahead energy/reserve demands to TSO. P, Q, and R in Fig.1 stand for active power, reactive power, and reserve, respectively. Accordingly, the TSO adopts the RUC model to optimize energy and reserve schemes of TGs with the lowest cost in the worst-case scenario, and provides the unit commitment (UC) status and worst-case uncertainty realization of load and wind power for robust economic dispatch (RED). Then, the Lagrange function is obtained from the RED model to derive LMP and ULMP. The LMP is employed to price energy. The ULMP is introduced to charge uncertainties and reward reserve provision.

In the lower level, each DSO conducts DS scheduling in Stage 2 and DEM pricing in Stage 3. Various types of DERs, i.e., energy storage systems (ESSs), microturbines (MTs), and RDG bid into the DEM in this paper. In Stage 2, the DSO receives the day-ahead forecast value and forecast deviation from RDG, and the boundary LMP/ULMP from TSO. Each DSO establishes a RO model to schedule ESSs and MTs to provide energy and reserve. Meanwhile, the energy/reserve demands of each DS are determined and sent to TSO. Stage 1 and Stage 2 are calculated iteratively by the HGD algorithm until the boundary LMP/ULMP converge. Then, the Stage 3 carries out the DEM pricing based on the linearized power flow model to derive DLMP$^{P}$, DLMP$^{Q}$, and DLMP$^{U}$, each of which can be decomposed to marginal costs for energy, congestion, voltage, and loss, as well as reflect uncertainties. Accordingly, this novel DEM pricing mechanism can provide effective price signals to manage not only the congestion and voltage, but also the uncertainty in DSs, and reflect the WEM-DEM interaction.

## III. Day-Ahead WEM Clearing and Pricing

### A. RUC and RED Models

The RUC/RED model includes dispatch and redispatch processes. In the dispatch process, TGs provide active energy according to the forecast value of WFs/LSEs and the energy demand of DSs. In the redispatch process, reserve capacity is optimized to cope with the forecast deviation of WFs/LSEs and the reserve demand of DSs. The RUC model is formulated as:

$$\min_{\boldsymbol{u},\boldsymbol{v},\boldsymbol{I}} \sum_{t\in T}\sum_{i\in\mathcal{G}_\mathcal{T}} (C_i^{SU}u_{i,t}+C_i^{SD}v_{i,t}) + \max_{\boldsymbol{\delta}\in\boldsymbol{U_1}} \min_{(\boldsymbol{P^G},\Delta\boldsymbol{P^G})\in\Omega(\boldsymbol{u},\boldsymbol{v},\boldsymbol{I},\boldsymbol{\delta})} \sum_{t\in T}\sum_{i\in\mathcal{G}_\mathcal{T}} (C_i^G(P_{i,t}^G)+C_i^R(\Delta P_{i,t}^G)) \quad (1)$$

s.t.

$$\sum_{i\in\mathcal{G}_\mathcal{T}} P_{i,t}^G=\sum_{i\in\mathcal{D}_\mathcal{T}} P_{i,t}^D+\sum_{i\in\mathcal{DS}_\mathcal{T}} P_{i,t}^{DS}-\sum_{i\in\mathcal{W}_\mathcal{T}} P_{i,t}^W, \forall t\ (\lambda_t^B) \quad (2a)$$

$$I_{i,t}P_{i,\min}^G\le P_{i,t}^G\le I_{i,t}P_{i,\max}^G, \forall i\in\mathcal{G}_\mathcal{T},\forall t \quad (\beta_{i,t}^{B-},\beta_{i,t}^{B+}) \quad (2b)$$

$$-r_i^{RD}(1-v_{i,t})-r_i^{SD}v_{i,t}\le P_{i,t}^G-P_{i,t-1}^G \le r_i^{RU}(1-u_{i,t})+r_i^{SU}u_{i,t}, \forall i\in\mathcal{G}_\mathcal{T},\forall t\ (\alpha_{i,t}^{B-},\alpha_{i,t}^{B+}) \quad (2c)$$

$$-F_l\le \sum_{i\in\mathcal{N}_\mathcal{T}} GSF_{l,i}\Big(\sum_{j\in\mathcal{G}_\mathcal{T}(i)} P_{j,t}^G+\sum_{j\in\mathcal{W}_\mathcal{T}(i)} P_{j,t}^W-\sum_{j\in\mathcal{D}_\mathcal{T}(i)} P_{j,t}^D -\sum_{j\in\mathcal{DS}_\mathcal{T}(i)} P_{j,t}^{DS}\Big)\le F_l, \forall l,t\ (\eta_{l,t}^{B-},\eta_{l,t}^{B+}) \quad (2d)$$

$$\sum_{i\in\mathcal{G}_\mathcal{T}} \Delta P_{i,t}^G=\sum_{i\in\mathcal{D}_\mathcal{T}} \delta_{i,t}^D+\sum_{i\in\mathcal{DS}_\mathcal{T}} R_{i,t}^{DS}-\sum_{i\in\mathcal{W}_\mathcal{T}} \delta_{i,t}^W, \forall t\ (\lambda_t^R) \quad (3a)$$

$$I_{i,t}P_{i,min}^G\le P_{i,t}^G+\Delta P_{i,t}^G\le I_{i,t}P_{i,max}^G, \forall i\in\mathcal{G}_\mathcal{T},\forall t\ (\beta_{i,t}^{R-},\beta_{i,t}^{R+}) \quad (3b)$$

$$-r_i^{RD}(1-v_{i,t})-r_i^{SD}v_{i,t}\le(P_{i,t}^G+\Delta P_{i,t}^G)-(P_{i,t-1}^G+\Delta P_{i,t-1}^G) \le r_i^{RU}(1-u_{i,t})+r_i^{SU}u_{i,t}, \forall i\in\mathcal{G}_\mathcal{T},\forall t\ (\alpha_{i,t}^{R-},\alpha_{i,t}^{R+}) \quad (3c)$$

$$-F_l\le \sum_{i\in\mathcal{N}_\mathcal{T}} GSF_{l,i}\Big(\sum_{j\in\mathcal{G}_\mathcal{T}(i)} (P_{j,t}^G+\Delta P_{j,t}^G)+\sum_{j\in\mathcal{W}_\mathcal{T}(i)} (P_{j,t}^W+\delta_{j,t}^W)- \sum_{j\in\mathcal{D}_\mathcal{T}(i)} (P_{j,t}^D+\delta_{j,t}^D)-\sum_{j\in\mathcal{DS}_\mathcal{T}(i)} (P_{j,t}^{DS}+R_{j,t}^{DS})\Big)\le F_l, \forall l,t\ (\eta_{l,t}^{R-},\eta_{l,t}^{R+}) \quad (3d)$$

$$\sum_{q=t-UT_i+1}^{t} u_{i,q}\le I_{i,t}, \forall i\in\mathcal{G}_\mathcal{T}, t\in[UT_i,T] \quad (4a)$$

$$\sum_{q=t-DT_i+1}^{t} v_{i,q}\le 1-I_{i,t}, \forall i\in\mathcal{G}_\mathcal{T}, t\in[DT_i,T] \quad (4b)$$

$$u_{i,t}+v_{i,t}\le 1, u_{i,t}-v_{i,t}=I_{i,t}-I_{i,t-1}, u_{i,t},v_{i,t},I_{i,t}\in\{0,1\}, \forall i\in\mathcal{G}_\mathcal{T},\forall t \quad (4c)$$

$$\boldsymbol{U_1}=\left\{\boldsymbol{\delta}:\ \begin{array}{l} -u_{i,t}^W\le\delta_{i,t}^W\le u_{i,t}^W, \forall i\in\mathcal{W}_\mathcal{T},\forall t \\ -u_{i,t}^D\le\delta_{i,t}^D\le u_{i,t}^D, \forall i\in\mathcal{D}_\mathcal{T},\forall t \end{array}\right\} \quad (5)$$

where $l$ denotes the line index; $F_l$ is the capacity (MW) of line $l$; $GSF_{l,i}$ is the generation shift factor (GSF) of bus $i$ to line $l$; $\mathcal{G}_\mathcal{T}(i)$, $\mathcal{W}_\mathcal{T}(i)$, $\mathcal{D}_\mathcal{T}(i)$, and $\mathcal{DS}_\mathcal{T}(i)$ denote the set of TGs, WFs, LSEs, and DSs connected to bus $i$; $u_{i,t}^W$ and $u_{i,t}^D$ are the maximum forecast deviation of WF/LSE $i$ respectively.

For TG $i$, $C_i^G(\cdot)$, $C_i^R(\cdot)$, $C_i^{SU}$, and $C_i^{SD}$ are the costs of energy, reserve, startup, and shutdown, respectively; $P_{i,\min}^G$ and $P_{i,\max}^G$ denote the lower/upper output limits (MW), respectively; $r_i^{RU}$, $r_i^{RD}$, $r_i^{SU}$, and $r_i^{SD}$ denote the maximum ramp-up/down and startup/shutdown ramp rate (MW/h), respectively; $UT_i$ and $DT_i$ are minimum up/down time (h), respectively. $\Omega(\boldsymbol{u},\boldsymbol{v},\boldsymbol{I},\boldsymbol{\delta})$ is the feasible region of $(\boldsymbol{P^G},\Delta\boldsymbol{P^G})$ for a given set of $(\boldsymbol{u},\boldsymbol{v},\boldsymbol{I},\boldsymbol{\delta})$. The variables in brackets of (2)-(3) are dual variables.

(1) minimizes the total operation cost, where the outer "min" model optimizes UC status, and the inner "max-min" model minimizes the cost under the worst-case scenario. (2)-(4) are the constraints for dispatch process, redispatch process, and UC status, respectively. (2a) and (3a) represent the power balance constraints. (2b) and (3b) are the generation limits. (2c) and (3c) are the ramping constraints. (2d) and (3d) are the line capacity limits. (4a) and (4b) are the minimum up/down time constraints. (4c) determines the UC of TGs. (5) is the uncertainty set.

The RED model is a linear programming (LP) model as formulated in (6) by fixing the UC $(I_{i,t},u_{i,t},v_{i,t})$ and worst-case uncertainty realization $(\delta_{i,t}^W,\delta_{i,t}^D)$ obtained in the RUC model.

$$\min_{\boldsymbol{P^G},\Delta\boldsymbol{P^G}} \sum_{t\in T}\sum_{i\in\mathcal{G}_\mathcal{T}} (C_i^G(P_{i,t}^G)+C_i^R(\Delta P_{i,t}^G)) \quad (6)$$

$$\text{s.t. } (2)\text{-}(3)$$

### B. WEM Clearing and Pricing Mechanism

According to the definition, the LMP at bus $i$ is defined as the partial derivative of the Lagrange function obtained in the RED model with respect to the forecast value of net load considering the energy demand of DSs.

$$\pi_{i,t}^P=\partial\mathcal{L}/\partial\Big(\sum_{j\in\mathcal{D}_\mathcal{T}(i)} P_{j,t}^D+\sum_{j\in\mathcal{DS}_\mathcal{T}(i)} P_{j,t}^{DS}-\sum_{j\in\mathcal{W}_\mathcal{T}(i)} P_{j,t}^W\Big) =\lambda_t^B-\sum_l GSF_{l,i}(\eta_{l,t}^{B+}-\eta_{l,t}^{B-}+\eta_{l,t}^{R+}-\eta_{l,t}^{R-}) \quad (7)$$

The ULMP at bus $i$ is defined as the partial derivative of the Lagrange function with respect to the forecast deviation of net load considering the reserve demand of DSs.

$$\pi_{i,t}^U=\partial\mathcal{L}/\partial\Big(\sum_{j\in\mathcal{D}_\mathcal{T}(i)} \delta_{j,t}^D+\sum_{j\in\mathcal{DS}_\mathcal{T}(i)} R_{j,t}^{DS}-\sum_{j\in\mathcal{W}_\mathcal{T}(i)} \delta_{j,t}^W\Big) =\lambda_t^R-\sum_l GSF_{l,i}(\eta_{l,t}^{R+}-\eta_{l,t}^{R-}) \quad (8)$$

The redispatch process against uncertainties and DS reserve demand is demonstrated in (3), whose dual variables are included in (7) and (8). Thus, LMP and ULMP can reflect the system uncertainty and DS reserve demand.

Based on the definition, LMP and ULMP are employed to price energy and uncertainty/reserve, respectively. For a market participant *i*, it is assumed that the active power generation, reserve capacity, and uncertainty (forecast deviation) are $P_{i,t}^{TS}$, $R_{i,t}^{TS}$, and $U_{i,t}^{TS}$ at time *t*. In the WEM clearing mechanism, the revenue of the participant *i* at bus *m* and time *t* is

$$REV_{i,t}^{TS}=\pi_{m,t}^{P}P_{i,t}^{TS}+\pi_{m,t}^{U}R_{i,t}^{TS}-\pi_{m,t}^{U}U_{i,t}^{TS} \quad (9)$$

The revenue is related to energy, reserve, and uncertainty. The uncertainty increases the profit loss of uncertainty sources and system operation costs. This WEM clearing and pricing mechanism internalizes uncertainties, and can provide effective price signals for uncertainty management in WEMs.

# IV. Day-Ahead DEM Clearing and Pricing

This section presents the day-ahead DEM clearing model and derives the uncertainty-aware DLMPs. Similar to the WEM model, the DEM model also includes two steps. The first step develops scheduling schemes considering uncertainties and TS-DS coordination by a RO model. When the discrete variables of capacitor banks (CBs) and on load tap changer (OLTC) as well as the worst-case scenario are determined in the first step, an LP model is formulated in the second step to derive DLMPs. These two steps are analogous to the RUC and RED models in the upper level, respectively. In addition, each step also includes dispatch and redispatch processes to cope with forecast value and forecast deviation (uncertainty) in DSs, respectively.

## A. Market Participant and Volt/VAR Control Models

### 1) RDG

$$-P_{i,t}^{rdg}\frac{\sqrt{1-\kappa^2}}{\kappa}\le Q_{i,t}^{rdg}\le P_{i,t}^{rdg}\frac{\sqrt{1-\kappa^2}}{\kappa},\forall i\in\mathcal{RDG},\forall t \quad (10a)$$

$$-(P_{i,t}^{rdg}+\epsilon_{i,t}^{rdg})\frac{\sqrt{1-\kappa^2}}{\kappa}\le Q_{i,t}^{rdg}\le(P_{i,t}^{rdg}+\epsilon_{i,t}^{rdg})\frac{\sqrt{1-\kappa^2}}{\kappa},\forall i\in\mathcal{RDG},\forall t \quad (10b)$$

$$\boldsymbol{U}_2=\left\{\boldsymbol{\epsilon}:-u_{i,t}^{rdg}\le\epsilon_{i,t}^{rdg}\le u_{i,t}^{rdg},\forall i\in\mathcal{RDG},\forall t\right\} \quad (10c)$$

where $Q_{i,t}^{rdg}$ denotes the reactive power of RDG *i* at time *t*; $\kappa$ is the minimum power factor (set to 0.95); $u_{i,t}^{rdg}$ is the uncertainty bound in the uncertainty set $\boldsymbol{U}_2$.

(10a) and (10b) represent the reactive power constraints in dispatch and redispatch processes, respectively. (10c) denotes the uncertainty set for RDG forecast.

### 2) MTs

$$(P_{i,t}^{mt}+\Delta P_{i,t}^{mt})^2+(Q_{i,t}^{mt})^2\le(S_{i,\max}^{mt})^2,\forall i\in\mathcal{MT},\forall t \quad (11a)$$

$$-r_i^{RD}\le P_{i,t}^{mt}-P_{i,t-1}^{mt}\le r_i^{RU},\forall i\in\mathcal{MT},\forall t \quad (11b)$$

$$-r_i^{RD}\le(P_{i,t}^{mt}+\Delta P_{i,t}^{mt})-(P_{i,t-1}^{mt}+\Delta P_{i,t-1}^{mt})\le r_i^{RU},\forall i\in\mathcal{MT},\forall t \quad (11c)$$

where $P_{i,t}^{mt}$, $Q_{i,t}^{mt}$ , and $\Delta P_{i,t}^{mt}$ denote the active/reactive power and reserve provided by MT *i*; $S_{i,\max}^{mt}$ is the generation capacity.

(11a) is the generation capacity limit. (11b) and (11c) are the ramping limits in the dispatch/redispatch processes respectively.

### 3) ESSs

$$0\le P_{i,t}^{ess,c}\le P_{i,\max}^{ess,c},0\le P_{i,t}^{ess,d}\le P_{i,\max}^{ess,d},\forall i\in\mathcal{ESS},\forall t \quad (12a)$$

$$E_{i,t}=E_{i,t\text{-}1}+\eta_i^c P_{i,t}^{ess,c}-(1/\eta_i^d)P_{i,t}^{ess,d},\forall i\in\mathcal{ESS},\forall t \quad (12b)$$

$$E_i^r SOC_{i,\min}\le E_{i,t}\le E_i^r SOC_{i,\max},\forall i\in\mathcal{ESS},\forall t \quad (12c)$$

$$E_{i,t=T}=E_{i,t=0},\forall i\in\mathcal{ESS},\forall t \quad (12d)$$

$$(P_{i,t}^{ess,c}-P_{i,t}^{ess,d})^2+(Q_{i,t}^{ess})^2\le(S_{i,\max}^{ess})^2,\forall i\in\mathcal{ESS},\forall t \quad (12e)$$

where $P_{i,t}^{ess,c}$, $P_{i,t}^{ess,d}$, and $Q_{i,t}^{ess}$ denote the charge/discharge active power and reactive power; $\eta_i^c$ and $\eta_i^d$ are the charge/discharge efficiency; $SOC_{i,\min}$ and $SOC_{i,\max}$ are the lower/upper limits of state of charge (SOC); $E_{i,t}$, $E_i^r$ and $S_{i,\max}^{ess}$ are remaining energy, rated energy capacity, and inverter capacity, respectively.

(12a) are the charge and discharge power constraints. (12b) reveals the relationship of remaining energy between adjacent periods. (12c) denotes the SOC limits. (12d) indicates that the stored energy at the beginning is equal to that at the end of a day. (12e) is the inverter capacity limit.

### 4) Volt/VAR Control

The Volt/VAR control is implemented by adjusting CBs, OLTC, and static VAR compensators (SVCs), which should meet the following constraints.

$$Q_{i,\min}^{svc}\le Q_{i,t}^{svc}\le Q_{i,\max}^{svc},\forall i\in\mathcal{SVC},\forall t \quad (13)$$

$$Q_{i,t}^{cb}=N_{i,t}^{cb}Q_{i,step}^{cb},\ N_{i,t}^{cb}\le N_{i,\max}^{cb},\forall i\in\mathcal{CB},\forall t \quad (14a)$$

$$\left|Q_{i,t}^{cb}-Q_{i,t-1}^{cb}\right|\ge Q_{i,t}^{cb}-Q_{i,t-1}^{cb},\left|Q_{i,t}^{cb}-Q_{i,t-1}^{cb}\right|\ge Q_{i,t-1}^{cb}-Q_{i,t}^{cb},\forall i\in\mathcal{CB},\forall t \quad (14b)$$

$$\left|Q_{i,t}^{cb}-Q_{i,t-1}^{cb}\right|\le\phi_{i,t}N_{i,\max}^{cb}Q_{i,step}^{cb},\sum_{t\in T}\phi_{i,t}\le N_{cap},\forall i\in\mathcal{CB},\forall t \quad (14c)$$

$$V_{1,t}^2=\sum_{k=0}^{K}(V_{\min}+\alpha\cdot k)^2 o_{k,t},\sum_{k=0}^{K}o_{k,t}=1,\forall t \quad (15a)$$

$$m_{k,t}\ge o_{k,t}-o_{k,t-1},m_{k,t}\ge o_{k,t-1}-o_{k,t},\psi_t=\frac{1}{2}\sum_{k=0}^{K}m_{k,t},\forall k,\forall t \quad (15b)$$

$$\sum_{t\in T}\psi_t\le N_{tp} \quad (15c)$$

where $Q_{i,t}^{svc}$ and $Q_{i,t}^{cb}$ denote the reactive power of SVC/CB *i*, respectively; $Q_{i,\min}^{svc}$ and $Q_{i,\max}^{svc}$ are the lower/upper capacity limits of SVC *i*; $N_{i,\max}^{cb}$ is the unit number in CB *i*; $Q_{i,step}^{cb}$ is the capacity of each unit in CB *i*. For the OLTC, $V_{1,t}$ is the voltage at the root node; *k* is the tap position; $\alpha$ is the voltage change per step; *K* is the total steps; $m_{k,t}$ is a binary variable indicating the action of the *k*th tap position, $m_{k,t}=1$ if the *k*th tap position changes, 0 otherwise. $N_{cap}$ and $N_{tp}$ are the maximum action number of a CB and OLTC, respectively.

(13) denotes the reactive power limit of SVCs. (14a) is the reactive power and the upper limit of CBs. (14b)-(14c) denote the linear constraints of action number for CBs. (15a) denotes the voltage model of OLTC. (15b) and (15c) represent the linear constraints of action number for the OLTC.

## B. Day-Ahead Robust Scheduling for Each DS

The day-ahead robust scheduling model for a DS at bus *m* in TS can be formulated as follows.

$$\min_{N_i^{cb},o_k}\sum_{t\in T}\left(\sum_{i\in\mathcal{CB}}c_i^c\phi_{i,t}+c^o\psi_t\right)$$

$$+\max_{\boldsymbol{\epsilon}\in\boldsymbol{U}_2}\ \min_{((\boldsymbol{P},\boldsymbol{Q},\boldsymbol{R})_m^{DS},(\boldsymbol{P},\boldsymbol{Q})_i^g,\Delta\boldsymbol{P}_i^{mt})\in\Omega(\boldsymbol{N}_i^{cb},\boldsymbol{o}_k,\boldsymbol{\epsilon})}\sum_{t\in T}(\pi_{m,t}^{P}P_{m,t}^{DS}+\pi_{m,t}^{Q}Q_{m,t}^{DS}$$

$$+\pi_{m,t}^{U}R_{m,t}^{DS}+\sum_{i\in\mathcal{DER}}(c_i^pP_{i,t}^g+c_i^q\widehat{Q}_{i,t}^g)+\sum_{i\in\mathcal{MT}}c_i^r\Delta P_{i,t}^{mt}) \quad (16)$$

$$\mathcal{DER}=\{\mathcal{RDG},\mathcal{MT},\mathcal{ESS}\}$$

s.t. (10)-(15)

$$u_{i,t}^b=(V_{i,t}^b)^2,w_{ij,t}^b=(I_{ij,t}^b)^2,\forall i\in\mathcal{N},\forall(i,j)\in\mathcal{L},\forall t \quad (17a)$$

$$\sum_{k\in pr(i)}(P_{ki,t}^{f,b}-r_{ki}w_{ki,t}^b)-\sum_{j\in cr(i)}P_{ij,t}^{f,b}=P_{i,t}^d-P_{i,t}^g,\forall t \quad (17b)$$

$$\sum_{k\in pr(i)}(Q_{ki,t}^{f,b}-x_{ki}w_{ki,t}^b)-\sum_{j\in cr(i)}Q_{ij,t}^{f,b}=Q_{i,t}^d-Q_{i,t}^g-Q_{i,t}^{svc}-Q_{i,t}^{cb},\forall t \quad (17c)$$

$$u_{i,t}^b-2(r_{ij}P_{ij,t}^{f,b}+x_{ij}Q_{ij,t}^{f,b})+(r_{ij}^2+x_{ij}^2)w_{ij,t}^b=u_{j,t}^b,j\in cr(i),\forall i\in\mathcal{N},t \quad (17d)$$

$$V_{i,\min}^2\le u_{i,t}^b\le V_{i,\max}^2,\forall i\in\mathcal{N},\forall t \quad (17e)$$

$$0 \leq w^b_{ij,t} \leq I^2_{ij,\max}, \forall (i,j) \in \mathcal{L}, \forall t \tag{17f}$$

$$(P^{f,b}_{ij,t})^2 + (Q^{f,b}_{ij,t})^2 \leq (S^f_{ij,\max})^2, \forall (i,j) \in \mathcal{L}, \forall t \tag{17g}$$

$$\left\| \begin{bmatrix} 2P^{f,b}_{ij,t} & 2Q^{f,b}_{ij,t} & w^b_{ij,t} - u^b_{i,t} \end{bmatrix}^{\mathrm{T}} \right\|_2 \leq w^b_{ij,t} + u^b_{i,t}, \forall (i,j) \in \mathcal{L}, \forall t \tag{17h}$$

$$u^r_{i,t} = (V^r_{i,t})^2, w^r_{ij,t} = (I^r_{ij,t})^2, \forall i \in \mathcal{N}, \forall (i,j) \in \mathcal{L}, \forall t \tag{18a}$$

$$\sum_{k \in pr(i)} (P^{f,r}_{ki,t} - r_{ki} w^r_{ki,t}) - \sum_{j \in cr(i)} P^{f,r}_{ij,t} = P^d_{i,t} - P^g_{i,t} - \epsilon^{rdg}_{i,t} - \Delta P^{mt}_{i,t}, \forall t \tag{18b}$$

$$\sum_{k \in pr(i)} (Q^{f,r}_{ki,t} - x_{ki} w^r_{ki,t}) - \sum_{j \in cr(i)} Q^{f,r}_{ij,t} = Q^d_{i,t} - Q^g_{i,t} - Q^{svc}_{i,t} - Q^{cb}_{i,t}, \forall t \tag{18c}$$

$$u^r_{i,t} - 2(r_{ij} P^{f,r}_{ij,t} + x_{ij} Q^{f,r}_{ij,t}) + (r^2_{ij} + x^2_{ij}) w^r_{ij,t} = u^r_{j,t}, j \in cr(i), \forall i \in \mathcal{N}, t \tag{18d}$$

$$V^2_{i,\min} \leq u^r_{i,t} \leq V^2_{i,\max}, \forall i \in \mathcal{N}, \forall t \tag{18e}$$

$$0 \leq w^r_{ij,t} \leq I^2_{ij,\max}, \forall (i,j) \in \mathcal{L}, \forall t \tag{18f}$$

$$(P^{f,r}_{ij,t})^2 + (Q^{f,r}_{ij,t})^2 \leq (S^f_{ij,\max})^2, \forall (i,j) \in \mathcal{L}, \forall t \tag{18g}$$

$$\left\| \begin{bmatrix} 2P^{f,r}_{ij,t} & 2Q^{f,r}_{ij,t} & w^r_{ij,t} - u^r_{i,t} \end{bmatrix}^{\mathrm{T}} \right\|_2 \leq w^r_{ij,t} + u^r_{i,t}, \forall (i,j) \in \mathcal{L}, \forall t \tag{18h}$$

$$\widehat{Q}^g_{i,t} \geq Q^g_{i,t}, \widehat{Q}^g_{i,t} \geq -Q^g_{i,t}, \forall i \in \mathcal{N}, \forall t \tag{19}$$

where $c^c_i$ and $c^o$ denote the action cost of CBs and OLTCs per time respectively; $\pi^Q_{m,t}$ denotes the reactive power LMP at bus $m$ in TS; $c^p_i$ and $c^q_i$ are active/reactive power bid prices of DERs; $\widehat{Q}^g_{i,t}$ is the absolute value of $Q^g_{i,t}$; $c^r_i$ and $\Delta P^{mt}_{i,t}$ are the reserve bid price and reserve capacity provided by MT $i$; $r_{ij}$ and $x_{ij}$ are the resistance and reactance of line $ij$; $V_{i,\min}$ and $V_{i,\max}$ are the minimum/maximum nodal voltage limits; $I_{ij,\max}$ is the current limit of line $ij$; $S_{ij,\max}$ is the capacity of line $ij$; $V_{i,t}$ is the voltage of node $i$; $I_{ij,t}$ is the current of line $ij$; $u_{i,t}$ and $w_{ij,t}$ denote the square values of $V_{i,t}$ and $I_{ij,t}$ respectively; $P^f_{ij,t}$ and $Q^f_{ij,t}$ are active/reactive power flow on line $ij$; the superscripts "$b$" and "$r$" correspond to dispatch/redispatch processes respectively.

(16) minimizes the total costs of regulating CBs and OLTC, purchasing active/reactive energy and reserve from the WEM, and scheduling the active/reactive power of DERs and reserve of MTs, respectively. The outer "min" model calculates the discrete variables, while the inner "max-min" model determines the continuous variables and worst-case uncertainty realization.

(17) and (18) are constraints for dispatch and redispatch processes, respectively. (17b)-(17c) and (18b)-(18c) are nodal active/reactive power balance constraints. (17d) and (18d) are nodal voltage equations. (17e)-(17f) and (18e)-(18f) indicate voltage and current limits. (17g) and (18g) are line power flow limits. (17h) and (18h) is a second-order-cone form for the relationship between line power flow, current and voltage. (19) linearizes the reactive power of DERs.

### *C. DEM Clearing and Pricing Mechanism*

After solving the scheduling model (10)-(19), the worst-case uncertainty realization and discrete variables can be determined and will be fixed to establish an LP model for DEM pricing based on the following linearized power flow model.

*1) Polygonal Inner-Approximation Method*

A polygonal inner-approximation method is utilized to linearize the quadratic capacity constraints (11a), (12e), (17g), and (18g), which can be formulated as follows [17]:

$$\alpha_{c,0} P + \alpha_{c,1} Q + \alpha_{c,2} S \leq 0, \forall c \in \{1,2,\ldots,12\} \tag{20}$$

where $\alpha_{c,0}$, $\alpha_{c,1}$, and $\alpha_{c,2}$ are the linearized coefficients.

*2) Branch Flow and Voltage Sensitivity Factors*

Branch flow and voltage can be expressed as [1]:

$$\boldsymbol{P^f} = \boldsymbol{LSF} \cdot \boldsymbol{P_{inj}}, \boldsymbol{Q^f} = \boldsymbol{LSF} \cdot \boldsymbol{Q_{inj}} \tag{21}$$

$$\Delta \boldsymbol{V} = \boldsymbol{R}\boldsymbol{P^f} + \boldsymbol{X}\boldsymbol{Q^f} = \boldsymbol{R} \cdot \boldsymbol{LSF} \cdot \boldsymbol{P_{inj}} + \boldsymbol{X} \cdot \boldsymbol{LSF} \cdot \boldsymbol{Q_{inj}} \tag{22}$$

where $\boldsymbol{P_{inj}}$ and $\boldsymbol{Q_{inj}}$ are nodal active/reactive power injections; $\Delta\boldsymbol{V}$ is the voltage change relative to the root node; $\boldsymbol{R}$ and $\boldsymbol{X}$ are the resistance/reactance matrix between any two nodes on the path; $\boldsymbol{LSF}$ is the load shift factor indicating the ratio of line loading change with respect to the injected nodal load.

The sensitivity factors (SFs) are derived from (21) and (22).

$$\boldsymbol{SF_{vp}} = \frac{\partial \Delta \boldsymbol{V}}{\partial \boldsymbol{P_{inj}}} = \boldsymbol{R} \cdot \boldsymbol{LSF}, \boldsymbol{SF_{vq}} = \frac{\partial \Delta \boldsymbol{V}}{\partial \boldsymbol{Q_{inj}}} = \boldsymbol{X} \cdot \boldsymbol{LSF} \tag{23}$$

$$\boldsymbol{SF_{lp}} = \frac{\partial \boldsymbol{P^f}}{\partial \boldsymbol{P_{inj}}} = \boldsymbol{LSF}, \boldsymbol{SF_{lq}} = \frac{\partial \boldsymbol{Q^f}}{\partial \boldsymbol{Q_{inj}}} = \boldsymbol{LSF} \tag{24}$$

where $\boldsymbol{SF_{vp}}$ and $\boldsymbol{SF_{vq}}$ are the SFs of nodal voltage change with respect to nodal active/reactive power injections; $\boldsymbol{SF_{lp}}$ and $\boldsymbol{SF_{lq}}$ are the SFs of line flows with respect to nodal active/reactive power injections.

*3) Delivery Factor and Fictitious Nodal Demand*

To account for the network loss in DSs, the delivery factor (DF) and fictitious nodal demand (FND) are utilized in the linearized power flow model. DF is a ratio representing the part of injected power that can be actually transmitted in the network considering the losses. Based on the DS scheduling results, the DF for active/reactive power at node $i$ can be expressed as [1]:

$$DF^p_{i,t} = 1 - \sum_{l \in \mathcal{L}} 2P^{f*}_{l,t} r_l \sum_{i \in \mathcal{N}} SF_{lp,l-i} \tag{25}$$

$$DF^q_{i,t} = 1 - \sum_{l \in \mathcal{L}} 2Q^{f*}_{l,t} x_l \sum_{i \in \mathcal{N}} SF_{lq,l-i} \tag{26}$$

where $P^{f*}_{l,t}$ and $Q^{f*}_{l,t}$ are the active/reactive power flow of line $l$ in the DS scheduling results. Note that all the variables with * represent the fixed values obtained in the DS scheduling model.

FND is a virtual nodal demand to stand for network losses which assigns the power loss of a line to its nodes at both sides equally. The FND for active power and reactive power can be expressed as:

$$F^p_{i,t} = \frac{1}{2} \sum_{j \in \mathcal{N}(i)} r_{ij} (P^{f*}_{ij,t})^2, F^q_{i,t} = \frac{1}{2} \sum_{j \in \mathcal{N}(i)} x_{ij} (Q^{f*}_{ij,t})^2 \tag{27}$$

*4) DEM Clearing and Pricing Mechanism*

The LP model for DEM pricing can be formulated as:

$$\min C(\boldsymbol{x}^*, \boldsymbol{\epsilon}^*, \boldsymbol{y}) \tag{28}$$

$$\sum_i DF^{p,b}_{i,t} P^g_{i,t} - \sum_i DF^{p,b}_{i,t} P^d_{i,t} - P^{b*}_{loss} = 0, \forall i \in \mathcal{N}, t \quad (\lambda^b_{p,t}) \tag{29a}$$

$$\sum_i DF^{q,b}_{i,t} Q^g_{i,t} - \sum_i DF^{q,b}_{i,t} Q^d_{i,t} - Q^{b*}_{loss} = 0, \forall i \in \mathcal{N}, t \quad (\lambda^b_{q,t}) \tag{29b}$$

$$P^g_{i,\min} \leq P^g_{i,t} \leq P^g_{i,\max}, \forall i \in \mathcal{N}, t \quad (\beta^{b-}_{p,i,t}, \beta^{b+}_{p,i,t}) \tag{29c}$$

$$Q^g_{i,\min} \leq Q^g_{i,t} \leq Q^g_{i,\max}, \forall i \in \mathcal{N}, t \quad (\beta^{b-}_{q,i,t}, \beta^{b+}_{q,i,t}) \tag{29d}$$

$$V_{\min} \leq V_{1,t} - \sum_j SF_{vp,i-j,t} (P^d_{j,t} - P^g_{i,t} + F^{p,b}_{i,t}) - \sum_j SF_{vq,i-j,t} (Q^d_{j,t} - Q^g_{i,t} + F^{q,b}_{i,t}) \leq V_{\max}, \forall t \quad (\mu^{b-}_{i,t}, \mu^{b+}_{i,t}) \tag{29e}$$

$$\alpha_{c,0} \sum_i SF_{lp,l-i,t} (P^d_{j,t} - P^g_{i,t} + F^p_{i,t}) + \alpha_{c,1} \sum_i SF_{lq,l-i,t} (Q^d_{j,t} - Q^g_{i,t} + F^q_{i,t}) + \alpha_{c,2} S^f_{l,\max} \leq 0, \forall i \in \mathcal{N}, l \in \mathcal{L}, t, c \quad (\eta^b_{l,c,t}) \tag{29f}$$

$$\sum_i DF^{p,r}_{i,t} (\Delta P^g_{i,t} + \epsilon^{rdg}_{i,t}) - (P^{r*}_{loss} - P^{b*}_{loss}) = 0, \forall i \in \mathcal{N}, t \quad (\lambda^r_{p,t}) \tag{30a}$$

$$P^g_{i,\min} \leq P^g_{i,t} + \Delta P^g_{i,t} \leq P^g_{i,\max}, \forall i \in \mathcal{N}, t \quad (\beta^{r-}_{p,i,t}, \beta^{r+}_{p,i,t}) \tag{30b}$$

$$V_{\min} \leq V_{1,t} - \sum_j SF_{vp,i-j,t} (P^d_{i,t} - (P^g_{i,t} + \Delta P^g_{i,t}) - \epsilon^{rdg}_{i,t} + F^{p,r}_{i,t}) - \sum_j SF_{vq,i-j,t} (Q^d_{j,t} - Q^g_{i,t} + F^{q,r}_{i,t}) \leq V_{\max}, \forall t \quad (\mu^{r-}_{i,t}, \mu^{r+}_{i,t}) \tag{30c}$$

$$\alpha_{c,0}\sum_i SF_{lp,l-i,t}(P^d_{i,t}-(P^g_{i,t}+\Delta P^g_{i,t})-\epsilon^{rdg}_{i,t}+F^{p,r}_{i,t})+\alpha_{c,1}\sum_i SF_{lq,l-i,t}(Q^d_{j,t}-Q^g_{i,t}+F^{q,r}_{i,t})+\alpha_{c,2}S^f_{l,\max}\leq 0,\forall i\in\mathcal{N},l\in\mathcal{L},t,c \quad (\eta^r_{l,c,t}) \tag{30d}$$

where $C(\boldsymbol{x}^*,\boldsymbol{\epsilon}^*,\boldsymbol{y})$ corresponds to the objetive function (16), $\boldsymbol{x}$, $\boldsymbol{y}$, and $\boldsymbol{\epsilon}$ denote the discrete variables, continuous variables, and uncertainty set, respectively; $P^*_{loss}$ and $Q^*_{loss}$ are the total active/reactive power losses; the superscripts "$b$" and "$r$" correspond to the dispatch/redispatch processes respectively; the variables in brackets are dual variables.

(28) is the objective function. (29) and (30) are constraints for dispatch/redispatch processes, respectively. (29a), (29b), and (30a) are power balance constraints. (29c), (29d), and (30b) denote the generation limits. (29e) and (30c) represent voltage limits. (29f) and (30d) denote the linear capacity constraints.

The Lagrange function of the LP model can be obtained easily. Similar to the WEM pricing, the DLMP$^{P}$/DLMP$^{Q}$ at node $i$ are defined as the partial derivative of the Lagrange function with respect to the forecasted active/reactive load. The DLMP$^{U}$ is defined as the partial derivative of the Lagrange function with respect to the forecast deviation of net load at that node. The DLMP$^{P}$, DLMP$^{Q}$, and DLMP$^{U}$ are derived as follows.

$$\chi^p_{i,t}=\partial\mathcal{L}/\partial P^d_{i,t}=\lambda^b_{p,t}+\sum_j SF_{vp,i-j,t}(\mu^{b-}_{i,t}-\mu^{b+}_{i,t}+\mu^{r-}_{i,t}-\mu^{r+}_{i,t})+\sum_l\sum_c \alpha_{c,0}SF_{lp,l-i,t}(\eta^b_{l,c,t}+\eta^r_{l,c,t})+\lambda^b_{p,t}(DF^{p,b}_{i,t}-1) \tag{31}$$

$$\chi^q_{i,t}=\partial\mathcal{L}/\partial Q^d_{i,t}=\lambda^b_{q,t}+\sum_j SF_{vq,i-j,t}(\mu^{b-}_{i,t}-\mu^{b+}_{i,t}+\mu^{r-}_{i,t}-\mu^{r+}_{i,t})+\sum_l\sum_c \alpha_{c,1}SF_{lq,l-i,t}(\eta^b_{l,c,t}+\eta^r_{l,c,t})+\lambda^b_{q,t}(DF^{q,b}_{i,t}-1) \tag{32}$$

$$\chi^u_{i,t}=\partial\mathcal{L}/\partial(-\epsilon^{rdg}_{i,t})=\lambda^r_{p,t}+\sum_j SF_{vp,i-j,t}(\mu^{r-}_{i,t}-\mu^{r+}_{i,t})+\sum_l\sum_c \alpha_{c,0}SF_{lp,l-i,t}\eta^r_{l,c,t}+\lambda^r_{p,t}(DF^{p,r}_{i,t}-1) \tag{33}$$

Each DLMP contains energy, voltage, congestion, and loss components, which is consistent with the traditional DLMP method to ensure the practicability. In addition, each DLMP includes $\mu^{r+}_{i,t}/\mu^{r-}_{i,t}$ and $\eta^r_{l,c,t}$, which are the dual variables related to voltage and congestion in redispatch process (30) to deal with uncertainties. Thus, this DEM pricing mechanism internalize uncertainties to extend the application scope of traditional DLMP method to uncertainty management in DSs.

It can be found from (31)-(33) that different nodes in a DS have the same energy component for any of the three DLMPs. The diversity of each DLMP in a DS depends on the difference in voltage, congestion, and loss components for different nodes. According to the definition, DLMP$^{P}$/DLMP$^{Q}$/DLMP$^{U}$ are used to price active energy, reactive energy and uncertainty/reserve, respectively. The active/reactive power generation, reserve, and uncertainty (forecast deviation) of a participant $i$ at node $m$ and time $t$ in DEM are assumed as $P^{ds}_{i,t}$, $Q^{ds}_{i,t}$, $R^{ds}_{i,t}$, and $U^{ds}_{i,t}$, respectively. Its cleared revenue can be expressed as

$$rev^{ds}_{i,t}=\chi^p_{m,t}P^{ds}_{i,t}+\chi^q_{m,t}Q^{ds}_{i,t}+\chi^u_{m,t}R^{ds}_{i,t}-\chi^u_{m,t}U^{ds}_{i,t} \tag{34}$$

The revenue of a participant can be decomposed into terms not only for active/reactive power, reserve, and uncertainty, but also for energy, voltage, congestion, and loss. This DEM clearing and pricing mechanism accounts for the system uncertainties and provides different feasible decomposition ideas for the revenues of participants, which can guide the DS management clearly and efficiently.

## V. Solution Methodology

Each level of the bi-level model include two steps, where the first step conducts scheduling based on RO methods, and the second step is an LP model for market pricing. The RO model in the form of "min-max-min" solves the optimization problem in the worst-case scenario. The first "min" model determines the discrete variables. The following "max-min" model determines the continuous variables and worst-case uncertainty realization. The RO model can be solved effectively by the column-and-constraint generation (CCG) algorithm [18].

The information interaction and coordination optimization between TS and DSs for the bi-level model is realized by the HGD algorithm, which is demonstrated as follows.

### A. HGD Algorithm

The HGD algorithm presented in [13] decomposes the bi-level model into TS subproblem for TSO and DS subproblems for each DSO, which can be formulated as follows respectively:

$$\min_{\boldsymbol{x}_{TS}} \boldsymbol{c}_{TS}(\boldsymbol{x}_{TS}) \tag{35a}$$

$$\text{s.t.} \quad \boldsymbol{A}_{TS}\boldsymbol{x}_{TS}+\boldsymbol{A}_{TB}\boldsymbol{x}_{DB}=\boldsymbol{a}_{TS} \tag{35b}$$

$$\boldsymbol{B}_{TS}\boldsymbol{x}_{TS}+\boldsymbol{B}_{TB}\boldsymbol{x}_{DB}\geq\boldsymbol{b}_{TS} \tag{35c}$$

$$\boldsymbol{E}_{TS}\boldsymbol{x}_{TS}\geq\boldsymbol{e}_{TS} \tag{35d}$$

$$\min_{\boldsymbol{x}_{DS},\boldsymbol{x}_{DB}} \boldsymbol{c}_{DS}(\boldsymbol{x}_{DS})+\boldsymbol{\xi}^{\mathrm{T}}_{TB}\boldsymbol{x}_{DB} \tag{36a}$$

$$\text{s.t.} \quad \boldsymbol{A}_{DS}\boldsymbol{x}_{DS}+\boldsymbol{A}_{DB}\boldsymbol{x}_{DB}=\boldsymbol{a}_{DS} \tag{36b}$$

$$\boldsymbol{B}_{DS}\boldsymbol{x}_{DS}+\boldsymbol{B}_{DB}\boldsymbol{x}_{DB}\geq\boldsymbol{b}_{DS} \tag{36c}$$

$$\boldsymbol{E}_{DS}\boldsymbol{x}_{DS}\geq\boldsymbol{e}_{DS},\boldsymbol{E}_{DB}\boldsymbol{x}_{DB}\geq\boldsymbol{e}_{DB} \tag{36d}$$

where $\boldsymbol{x}_{TS}$ and $\boldsymbol{x}_{DS}$ are decision variables in the TS subproblem and DS subproblem respectively, $\boldsymbol{x}_{DB}$ is the boundary demands of DSs; $\boldsymbol{A}$, $\boldsymbol{B}$, and $\boldsymbol{E}$ are the coefficient matrices of operation constraints; $\boldsymbol{a}$, $\boldsymbol{b}$, and $\boldsymbol{e}$ are constant column vectors; the subscripts $TS$, $TB$, $DB$, and $DS$ of $\boldsymbol{A}$, $\boldsymbol{B}$, $\boldsymbol{E}$, and $\boldsymbol{e}$ indicate that the matrix/vector corresponds to $\boldsymbol{x}_{TS}$, $\boldsymbol{x}_{DB}$ in TS level, $\boldsymbol{x}_{DB}$ and $\boldsymbol{x}_{DS}$ in DS level, respectively; the subscripts $TS$ and $DS$ of $\boldsymbol{a}$ and $\boldsymbol{b}$ represent that the column vectors are in TS and DS levels, respectively; $\boldsymbol{\xi}_{TB}=-\boldsymbol{A}^{\mathrm{T}}_{TB}\boldsymbol{\lambda}_{\boldsymbol{a}_{\mathrm{T}}}-\boldsymbol{B}^{\mathrm{T}}_{TB}\boldsymbol{\omega}_{\boldsymbol{b}_{\mathrm{T}}}$, $\boldsymbol{\lambda}_{\boldsymbol{a}_{\mathrm{T}}}$ and $\boldsymbol{\omega}_{\boldsymbol{b}_{\mathrm{T}}}$ are the dual variables of constraints in (35), $\boldsymbol{\xi}_{TB}$ is verified as LMP in [13].

The RUC model (1)-(5) and the RO scheduling model (10)-(19) can be transformed to the TS subproblem (35) and DS subproblem (36), respectively. As stated in [13], $\boldsymbol{\xi}_{TB}$ stands for the boundary LMP/ULMP calculated by TSO, and $\boldsymbol{x}_{DB}$ is the energy/reserve demands of DSs in this paper. $\boldsymbol{\xi}_{TB}$ and $\boldsymbol{x}_{DB}$ are exchanged at the PSP, and the bi-level model can be solved iteratively by the HGD algorithm. The convergence criterion is

$$\max\left|\boldsymbol{\xi}_{TB,k}-\boldsymbol{\xi}_{TB,k-1}\right|\leq\varepsilon \tag{37}$$

where $k$ is the iteration number; $\varepsilon$ is the convergence tolerance.

However, the convergence of the bi-level optimization is a challenge. When the boundary prices in TS increase, the DS demands may decrease accordingly to reduce costs. In this condition, the boundary prices would fall back. That is, the boundary prices oscillate (rise-fall-rise) over the iterations and the DS demands will decrease-increase-decrease accordingly. This solution oscillation issue eventually leads to the divergence, which has been revealed in [13].

### B. LMP/ULMP-Sensitivity-Based HGD Algorithm

To ensure the convergence, an LMP/ULMP sensitivity-based HGD algorithm is proposed in this paper. It is a natural idea that, if the boundary LMP/ULMP ($\boldsymbol{\xi}_{TB}$) is evaluated by DSOs before DS scheduling in each iteration, the DS demands calculated in DS subproblems will not increase/decrease in the way to cause divergence. Inspired by [19], $\xi_{TB}$ can be evaluated by the sensitivity $\boldsymbol{S}_{\xi\text{-}\boldsymbol{x}_{DB}}$ of $\xi_{TB}$ to $\boldsymbol{x}_{DB}$, which is expressed as:

$$\tilde{\xi}_{TB,k+1}=\xi_{TB,k}+\boldsymbol{S}_{\xi\text{-}\boldsymbol{x}_{DB},k}(\boldsymbol{x}_{DB,k}-\boldsymbol{x}_{DB,k-1}) \tag{38}$$

where $\xi_{TB,k}$ is calculated by TSO; $\tilde{\xi}_{TB,k+1}$ is the estimated value.

A probing mechanism is utilized to calculate the sensitivity which is a response function of the previous boundary prices in TS and the DS demands. Each DSO records previous boundary data, and estimate $\boldsymbol{S}_{\xi\text{-}\boldsymbol{x}_{DB},k}$ after the third iteration in a distributed manner with light communication/computational cost, which is formulated as:

$$\boldsymbol{S}_{\xi\text{-}\boldsymbol{x}_{DB},k}=\left(\sum_{i=2}^{k-1}\left((\xi_{TB,i+1}-\xi_{TB,i})/(\boldsymbol{x}_{DB,i}-\boldsymbol{x}_{DB,i-1})\right)\right)/(k\text{-}2)\,,k\geq 3 \tag{39}$$

In (36a), replace $\xi_{TB}$ with $\tilde{\xi}_{TB,k}$ and multiply the coefficient of the last term in the expansion formulas by 0.5. Then the objective function for each DS subproblem is reformulated as:

$$\min_{\boldsymbol{x}_{DS},\boldsymbol{x}_{DB}} \boldsymbol{c}_{DS}(\boldsymbol{x}_{DS})+(\xi_{TB,k}-\boldsymbol{S}_{\xi\text{-}\boldsymbol{x}_{DB},k}\boldsymbol{x}_{DB,k-1})^{\mathrm{T}}\boldsymbol{x}_{DB}+\frac{1}{2}\boldsymbol{x}_{DB}^{\mathrm{T}}\boldsymbol{S}_{\xi\text{-}\boldsymbol{x}_{DB},k}\boldsymbol{x}_{DB} \tag{40}$$

The coefficient 0.5 of the last term in (40) guarantees the solution optimality [19]. The calculation flowchart of the bi-level model is shown in Fig. 2.

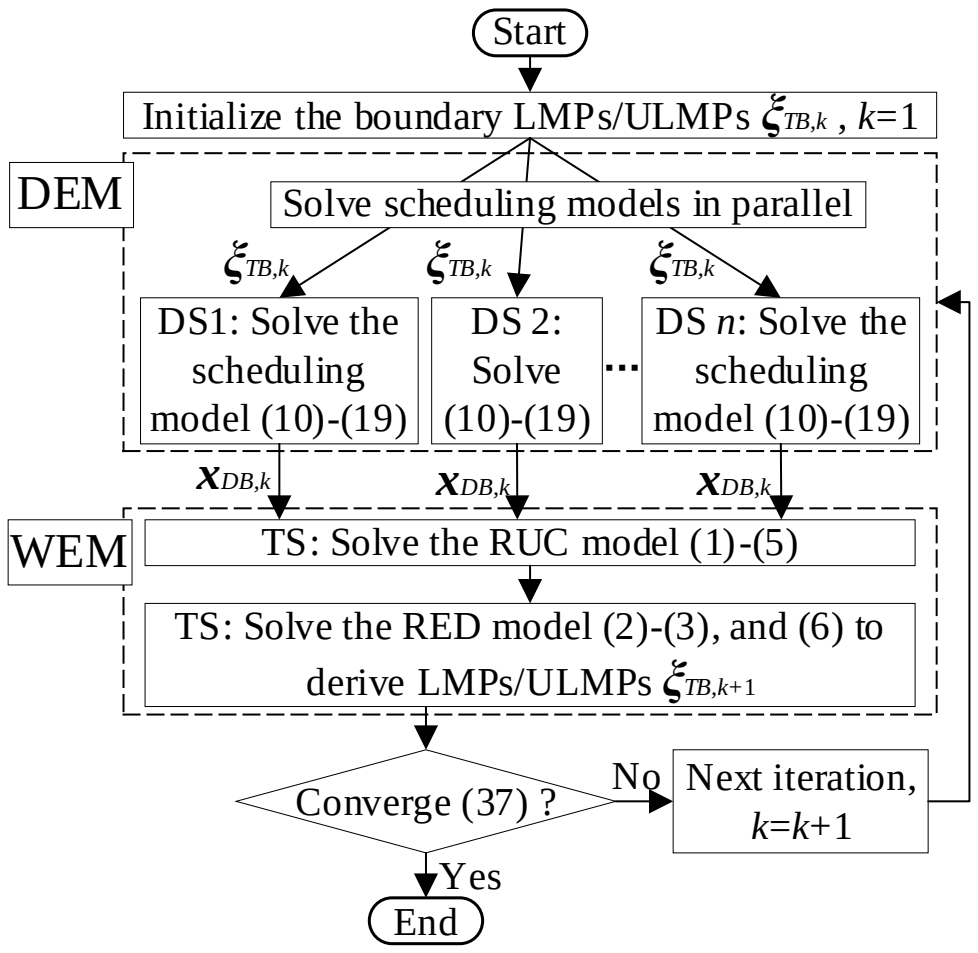

Fig.2. The calculation flowchart of the bi-level model.

## VI. Case Studies

The effectiveness of the proposed DEM clearing and pricing mechanism is verified by the T5D33 and T118D69 systems in this section. All simulations are performed in MATLAB 2020a with YALMIP interface and MOSEK 9.2 on a computer with Intel(R) Core (TM) i7-10700F CPU and 16 GB RAM.

### *A. Simulation Settings*

The T5D33 system is comprised of a modified PJM 5-bus TS depicted in Fig. 3 and 100 identical IEEE 33-node DSs, in which one half are at bus C and the other half are at bus D in the TS. Fig. 3 shows the capacity and bid prices of TGs. The other parameters are given in [20]. The ratio of the forecasted base load of LSEs at buses B, C, and D is 3:3:4. The maximum forecast deviation of the LSEs at buses B, C, and D is 10%, 5%, and 0% to their forecasts. A 200MW WF is located at bus E.

For the IEEE 33-node DS, the parameter configuration of load demands, line, SVCs, CBs, and OLTC are given in [1]. Each DS includes eight PVs with capacity of {0.6, 0.6, 0.5, 0.6, 0.5, 0.8, 0.6, 0.8} MW at nodes {4, 7, 11, 15, 18, 25, 28, 32}, two WTs with capacity of {0.2, 0.2} MW at nodes {13, 20}, two MTs with capacity of {0.8, 0.8} MW and ramp up/down rate limit of {0.4, 0.4} MW/h at nodes {17, 32}, two 3MW×2h ESSs at nodes {3, 29}. The voltage magnitude of the root node is set to 1.0 p.u.. The voltage limits are [0.95, 1.05] p.u..

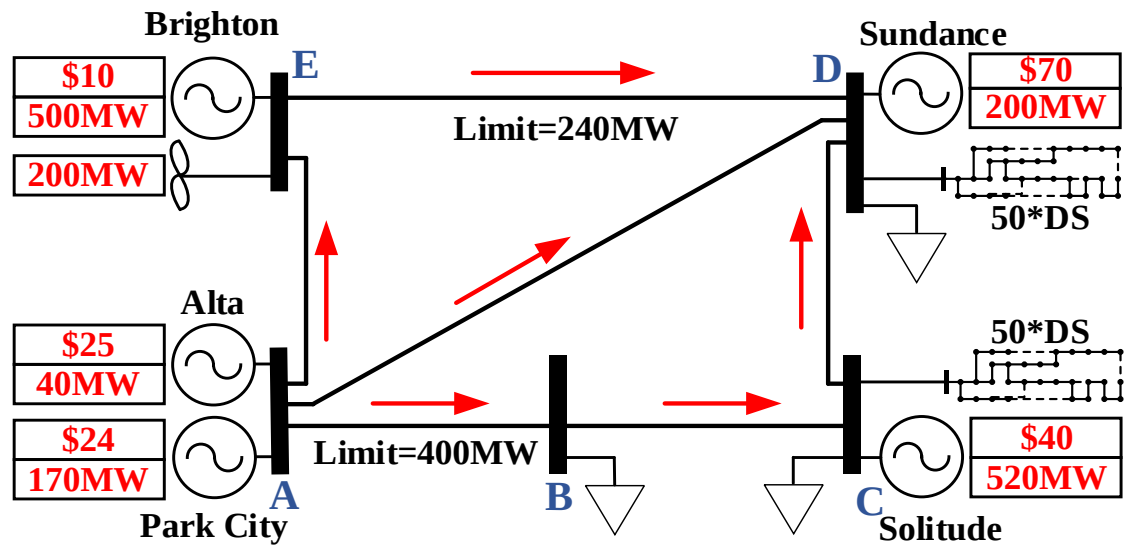

Fig.3. The modified PJM 5-bus system.

The WF in TS and the PVs/WTs in DSs follow the forecast profiles considering uncertainty in [21]. The MTs and ESSs bid at 15 \$/MWh and 20 \$/MWh for active power and 3 \$/MVArh and 4 \$/MVArh for reactive power, respectively. The TGs in TS and MTs in DSs provide reserve capacity at half of their active power bid prices [22]. The reactive power price in WEM is set to 10% LMP [1]. The LMPs for active and reactive power in WEM are denoted as $\text{LMP}^{\text{P}}$ and $\text{LMP}^{\text{Q}}$, respectively. The action costs of CBs and OLTCs are 0.24 \$/time and 1.40 \$/time, respectively [23]. The optimality gaps for each level of the bi-level model is set to 1%, as well as the HGD algorithm.

In order to verify the effectiveness of the proposed method, several cases are designed for comparison. Case 1 applies the above parameters settings and is regarded as a benchmark case. The parameter settings of other cases are as follows.

Case 2: There is no uncertainty of RDG in DSs.

Case 3: The maximum forecast deviation of RDG in DSs is set to twice that in Case 1.

Case 4: The DEM and WEM conduct clearing and pricing separately [1]. In the WEM, the boundary DS injection is regarded as a constant. The initial energy demand of a DS is equal to the load demand minus the forecasted power of RDG, regardless of the output of MTs and ESSs. The initial reserve demand of a DS is equal to the sum of worst-case forecast deviation of RDG, ignoring the reserve provision of MTs. Based on this boundary information, the TSO clears the WEM and sends the boundary LMP/ULMP to DSs. Accordingly, each DSO conducts self-scheduling and DEM pricing.

The other parameters in Cases 2-4 are the same with Case 1. Section VI-B/C illustrates the DEM/WEM clearing and pricing in Case 1. Section VI-D performs the sensitivity analysis on uncertainty by comparing Cases 1-3. Sections VI-E reveals the necessity of TS-DS coordination by comparing Cases 1 with 4.

### *B. WEM Clearing and Pricing*

The LMP/ULMP is shown in Fig. 4. Network congestion leads to different prices at different buses in hours 1-2, 7-8, and 17-24. The prices at bus D are the highest, followed by bus C.

The active energy and reserve demands of each DS at buses C and D are shown in Fig. 5. The “initial” curve denotes the net load of load demand, PVs, and WTs except GTs and ESSs. Fig. 5 (a) indicates that the active energy demand decreases under high LMP and vice versa to reduce costs, which is achieved by dispatching MTs and ESSs. Fig. 5 (b) reveals that the reserve demands of DSs decrease with the reserve provision from MTs.

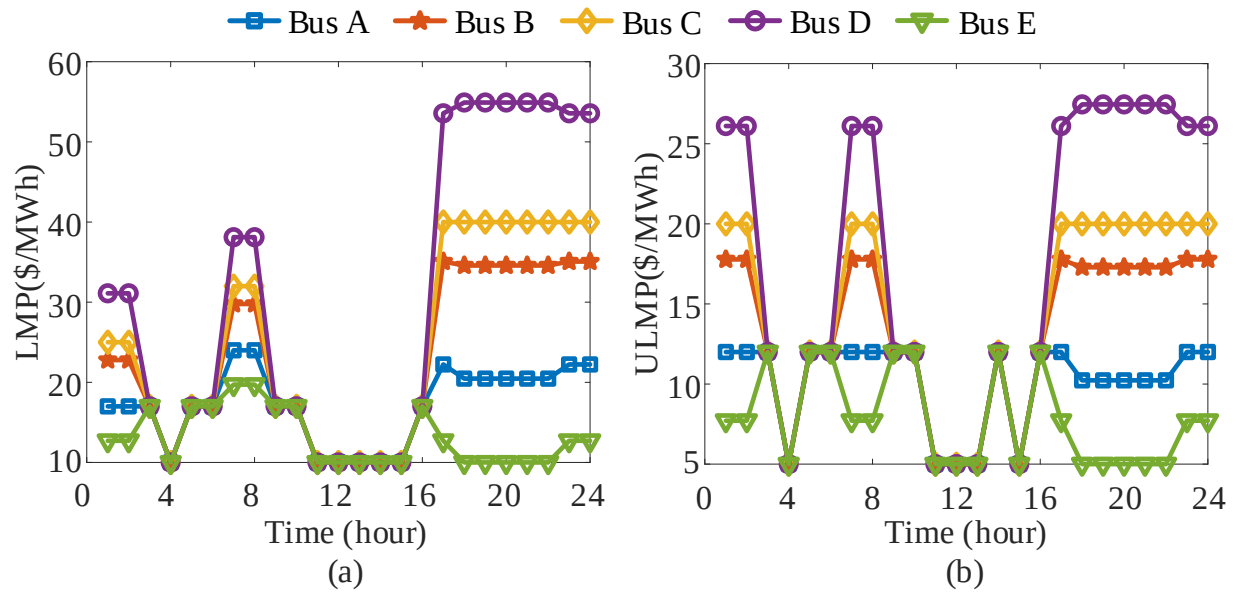

Fig.4. LMP and ULMP in WEM. (a) LMP. (b) ULMP.

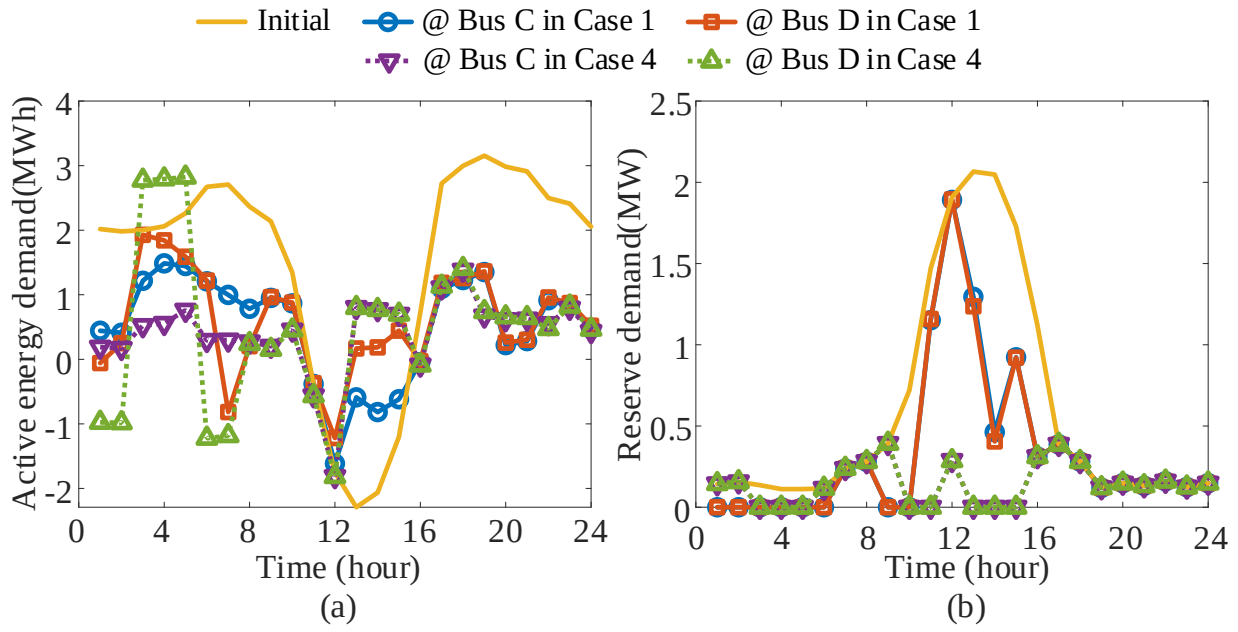

Fig.5. (a) Active energy and (b) reserve demands of each DS.

Table I shows the purchased energy/reserve of each DS from WEM and the charge/discharge energy of ESSs. Table II lists the operation costs of each DS. It can be seen that although bus D has a higher LMP than bus C, the DS at bus D purchases more active energy from WEM. The main reason is that the $LMP^P$ at bus D has a more significant price spread. Thus, the ESSs in the DS at bus D charge/discharge more active energy to reduce DS cost and gain more arbitrage. The loss in ESS charge/discharge process leads to a greater active energy demand from WEM to fill this gap. Because the $LMP^Q$ is lower at bus C and higher at bus D, the DSO at bus C purchases reactive energy and the DSO at bus D sells residual reactive energy from/to WEM. In terms of the reserve of DSs purchased from WEM, it is larger at bus C due to the lower ULMP. Due to the price differences in WEM, the DSO at bus D pays more costs on active energy and reserve, and can make a profit from WEM by selling reactive energy.

TABLE I
PURCHASE OF DSS FROM WEM AND CHARGE/DISCHARGE ENERGY OF ESSS

| Each DS | Active energy (MWh) | Reactive energy (MVarh) | Reserve (MW) | Charge energy (MWh) | Discharge energy (MWh) |
|---|---|---|---|---|---|
| @ Bus C | 12.02 | 6.37 | 8.05 | 3.49 | 2.82 |
| @ Bus D | 13.01 | -20.42 | 7.95 | 7.41 | 6.01 |

TABLE II
OPERATION COSTS OF EACH DS AT BUSES C AND D ($)

| Each DS | Active Power | Reactive Power | Reserve | Total |
|---|---|---|---|---|
| @ Bus C | 816.69 | 28.17 | 121.69 | 966.55 |
| @ Bus D | 925.50 | -1.61 | 135.02 | 1058.91 |

### C. *DEM Clearing and Pricing*

DLMPs in each DS are shown in Fig. 6-Fig. 8. It can be seen that DLMP profiles are closely related to the boundary prices which has been shown in Fig. 4. The DLMPs in the DSs at buses C and D are different due to the different location in TS.

Different nodes in a DS have the same energy component of DLMPs. From the DS perspective, TS is a virtual source whose marginal cost is LMP/ULMP. When DSs trade with TS and the traded energy/reserve is not congested by substations, TS can be regarded as the marginal unit in DSs. Thus, the boundary LMP/ULMP is equal to the energy component of corresponding DLMPs, which is the case in most hours in Case 1. However, in hours 1-3, 5-6, and 9-10, all DSs trade no reserve with TS. Their reserve demand is supplied by MTs, which determine the energy component of $DLMP^U$. In this situation, the boundary ULMP is not equal to the energy component of $DLMP^U$, which is shown in Table III by taking a DS at bus D as an example.

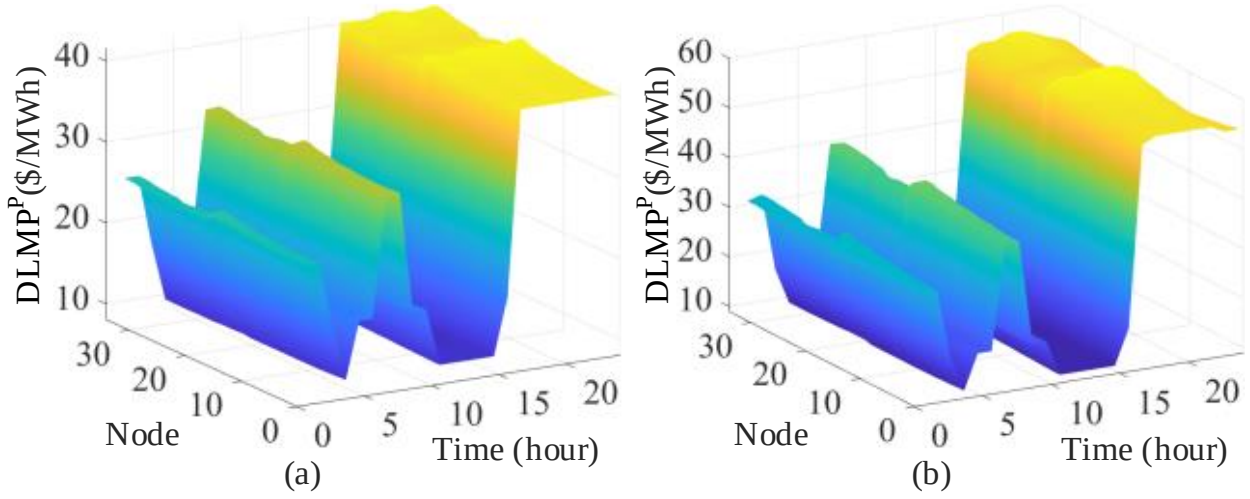

Fig.6. $DLMP^P$ in each DS at (a) bus C and (b) bus D.

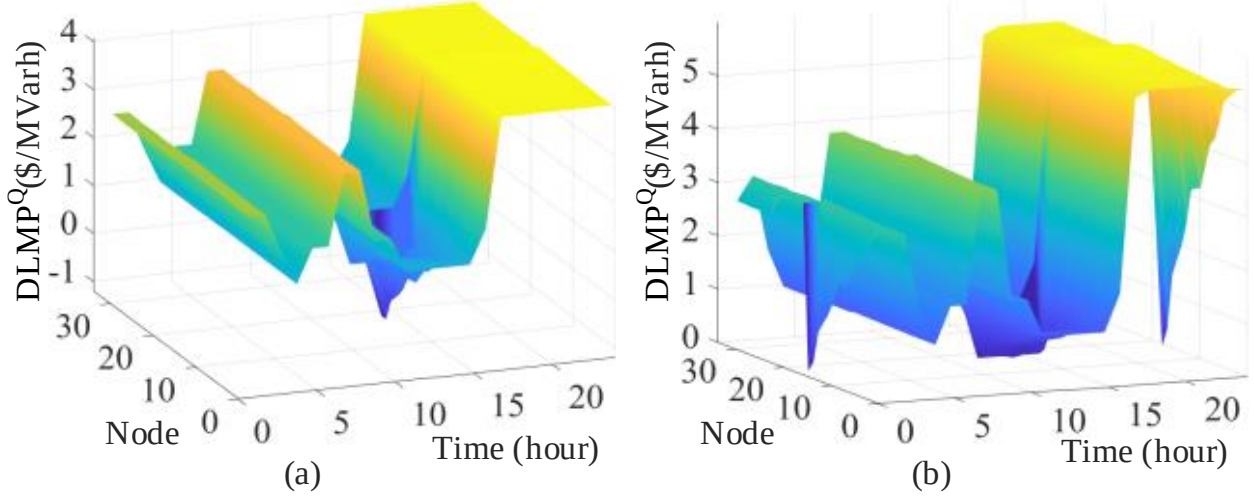

Fig.7. $DLMP^Q$ in each DS at (a) bus C and (b) bus D.

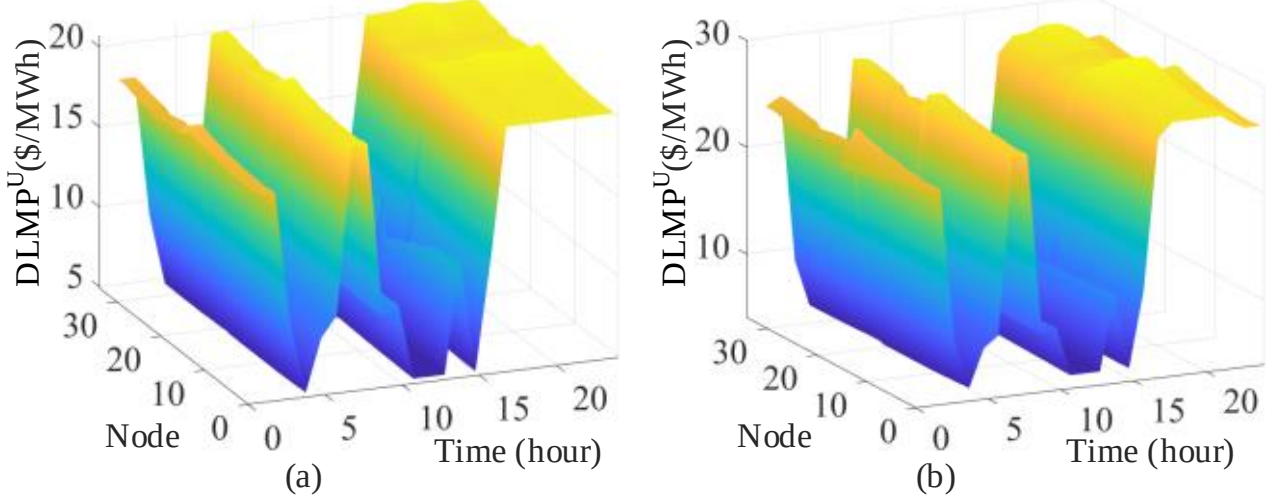

Fig.8. $DLMP^U$ in each DS at (a) bus C and (b) bus D.

TABLE III
ENERGY COMPONENT (E.COM.) OF $DLMP^U$ AND BOUNDARY ULMP ($/MWh)

| Hour | 1 | 2 | 3 | 5 | 6 | 9 | 10 |
|---|---|---|---|---|---|---|---|
| E.Com. | 23.81 | 23.91 | 9.26 | 8.51 | 9.52 | 9.20 | 9.54 |
| ULMP | 26.11 | 26.11 | 12.00 | 12.00 | 12.00 | 12.00 | 12.00 |

The diversity of $DLMP^P$/$DLMP^Q$/$DLMP^U$ in a DS depends on the difference in voltage, congestion, and loss components for different nodes, which provide price signals for managing voltage, congestion, and loss considering uncertainties. Taking node 18 in a DS at bus D as an example, the voltage/loss components in DLMPs and nodal voltage in certain hours are shown in Table IV. "V." denotes the voltage, "Ba." and "Re." denote dispatch/redispatch process in DS scheduling. The congestion component in DLMPs is zero in these hours.

TABLE IV
VOLTAGE/LOSS COMPONENTS IN DLMPS AND NODAL VOLTAGE AT NODE 18

| Hour | $DLMP^P$ | | | $DLMP^Q$ | | | $DLMP^U$ | | Voltage | |
|---|---|---|---|---|---|---|---|---|---|---|
| | V.Ba. | V.Re. | Loss | V.Ba. | V.Re. | Loss | V.Re. | Loss | Ba. | Re. |
| 1 | -0.33 | -3.55 | 2.02 | -0.29 | -2.93 | 0.12 | -3.55 | 1.20 | 1.05 | 1.05 |
| 4 | 0 | 0 | -0.40 | 0 | 0 | 0.03 | 0 | -0.18 | 0.98 | 0.99 |
| 13 | -1.20 | 0 | 1.07 | -1.00 | 0 | -0.01 | 0 | 0.16 | 1.05 | 1.03 |
| 14 | 0 | -1.20 | 1.02 | 0 | -1.00 | 0 | -1.20 | 0.65 | 1.03 | 1.05 |

It can be seen that overvoltage in dispatch/redispatch process leads the related voltage component to be nonzero. For instance, the voltage reaches the upper limit caused by the uncertainty of RDG in redispatch process in hour 14. Accordingly, the voltage components in redispatch process (V.Re.) are negative, which indicates that increasing nodal load injection at node 18 can reduce the DS cost on voltage and therefore alleviate voltage issue. In this way, the voltage component in DLMPs motivates market participants to increase/decrease output to implement voltage management while considering uncertainties. Similarly, the congestion component also reflects dispatch/redispatch processes, which can provide price signals for congestion management considering uncertainties. For the loss component in Table IV, the positive values indicate that decreasing the nodal load injection/uncertainty at node 18 can reduce the DS cost on loss and therefore reduce losses, and vice versa.

The above analysis reveals that uncertainty is internalized in DLMPs and their decomposition, which can provide effective price signals to DSO for managing voltage, congestion, and loss as well as the uncertainty.

## D. *Sensitivity Analysis of Uncertainty Level in DSs*

Table V shows the DEM results for each DS at bus D in TS, and Tables VI presents the WEM results, in Cases 1-3. It can be seen that high uncertainty in DSs increases not only the price directly related to uncertainty in DEM, i.e., $DLMP^U$, but also $DLMP^P$/$DLMP^Q$ and LMP/ULMP in both markets. The reason is that the uncertainty in DSs affects the overall reserve demand in TS, resulting in the change of ULMP. The coupling between uncertainty/reserve with active/reactive power and the TS-DS interaction lead to the change of LMP/$DLMP^P$/$DLMP^Q$. On the other hand, high uncertainty of RDG reduces their profit, and increases the profits of MTs and ESSs. Furthermore, high uncertainty and the increased reserve demand of DSs increase the reserve cost and total operation cost in both TS and DSs.

TABLE V
DEM CLEARING AND PRICING RESULTS OF EACH DS AT BUS D IN TS

| Case | | Case 1 | Case 2 | Case 3 |
|---|---|---|---|---|
| Average Price | $DLMP^P$ ($/MWh) | 30.77 | 30.33 | 37.50 |
| | $DLMP^Q$ ($/MVArh) | 2.94 | 2.89 | 3.57 |
| | $DLMP^U$ ($/MWh) | 17.16 | 0.00 | 22.05 |
| Revenue of PVs ($) | Active Power | 380.44 | 356.95 | 685.81 |
| | Reactive Power | 2.17 | 3.49 | 1.33 |
| | Reserve | -97.61 | 0.00 | -457.41 |
| | Profit | 284.99 | 360.45 | 229.74 |
| Revenue of WTs ($) | Active Power | 228.26 | 224.54 | 277.07 |
| | Reactive Power | 4.22 | 6.92 | 1.62 |
| | Reserve | -51.56 | 0.00 | -131.05 |
| | Profit | 180.91 | 231.46 | 147.64 |
| Revenue of MTs ($) | Active Power | 1028.64 | 1042.85 | 1187.07 |
| | Reactive Power | 0.00 | 0.00 | 0.00 |
| | Reserve | 59.39 | 0.00 | 226.75 |
| | Profit | 675.62 | 658.87 | 937.43 |
| Revenue of ESSs ($) | Active Power | 194.19 | 178.55 | 177.83 |
| | Reactive Power | 161.02 | 151.65 | 161.70 |
| | Profit | 111.34 | 106.94 | 115.62 |
| Purchase from WEM | Active Power (MWh) | 13.01 | 11.73 | 16.03 |
| | Reactive Power (MVArh) | -20.42 | -19.48 | -14.76 |
| | Reserve (MW) | 7.95 | 0.00 | 12.46 |
| Operation Cost of the DS ($) | Active Power | 925.50 | 923.78 | 981.50 |
| | Reactive Power | -1.61 | -9.98 | 3.27 |
| | Reserve | 135.02 | 0.00 | 393.03 |
| | Total | 1058.91 | 913.80 | 1377.79 |

The above results illustrate that the uncertainty in DSs affect not only DEM clearing and pricing, but also WEM clearing and pricing. High uncertainty of RDG reduces their profit, increases the operation cost and electricity prices in both TS and DSs. The proposed DLMP mechanism can stimulate uncertainty sources to improve forecast accuracy and provide effective price signals to incentivize the uncertainty management in DSs.

TABLE VI
WEM CLEARING AND PRICING RESULTS

| Case | | Case 1 | Case 2 | Case 3 |
|---|---|---|---|---|
| Average Price ($/MWh) | LMP | 21.71 | 21.42 | 25.03 |
| | ULMP | 12.75 | 11.48 | 15.26 |
| Operation Cost of the TS ($) | Energy | 206311.91 | 207058.77 | 207883.14 |
| | Reserve | 26891.83 | 16913.12 | 42423.13 |
| | Total | 233203.74 | 223971.88 | 250306.27 |

## E. *Necessity Analysis of WEM-DEM Coordination*

Fig. 5 also shows the initial/actual demand before/after self-scheduling for each DS. The large mismatch between the initial and actual demands indicates the power imbalance at the PSP, which argues that the separate scheduling is not feasible for the coordinated TS-DS system.

Fig. 9 shows the power flow of line D-E in TS in Cases 1 and 4. Congestion occurs 8/12 times in Case 1 and 14/16 times in Case 4 in dispatch/redispatch process, which indicates that congestion can be alleviated by TS-DS coordination.

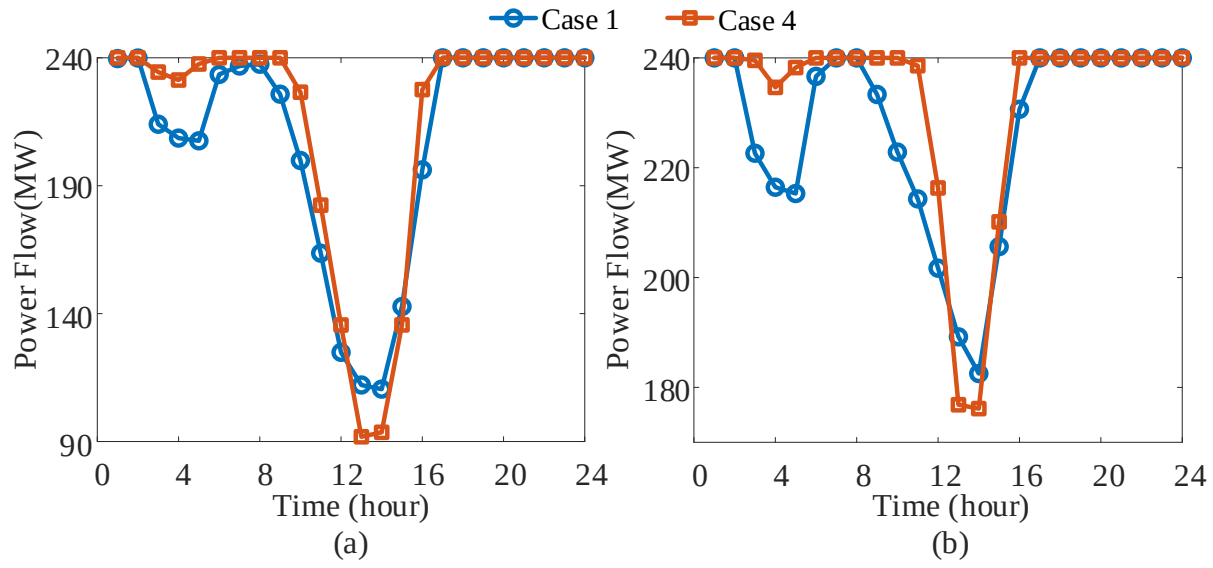

Fig.9. Power flow of line D-E in TS in (a) dispatch and (b) redispatch processes.

Table VII shows the average prices in Cases 1 and 4, which indicates that TS-DS coordination can effectively reduce the electricity prices in both WEM and DEM. Tables VIII lists the operation costs in TS and DS levels. Compared with Case 4, the total costs in TS and DS level are reduced by 3.38% and 29.39% in Case 1, which confirms that the TS-DS coordination can effectively reduce the overall operation costs.

TABLE VII
AVERAGE ELECTRICITY PRICES

| Case | LMP ($/MWh) | ULMP ($/MWh) | $DLMP^P$ ($/MWh) | $DLMP^Q$ ($/MVArh) | $DLMP^U$ ($/MWh) |
|---|---|---|---|---|---|
| Case 1 | 21.71 | 12.75 | 30.77 | 2.94 | 17.16 |
| Case 4 | 27.03 | 15.67 | 41.99 | 3.65 | 22.52 |

TABLE VIII
OPERATION COSTS IN TS LEVEL (WEM) AND DS LEVEL (DEM) ($)

| Case | DS level (DEM) | | | | TS level (WEM) | | |
|---|---|---|---|---|---|---|---|
| | Active | Reactive | Reserve | Total | Energy | Reserve | Total |
| Case 1 | 925.50 | -1.61 | 135.02 | 1058.91 | 206312 | 26892 | 233204 |
| Case 4 | 970.16 | -40.04 | 165.86 | 1095.98 | 294040 | 36087 | 330128 |

The above results verify the necessity and benefits of TS-DS coordination, which is achieved by iterative calculation between TS and DSs. The T5D33 system in Case 1 requires 6 iterations and 88.2 seconds. Taking the LMP and energy demand of a DS at bus D as an example, the convergency performance is shown in Fig.10. It can be seen that LMP is inversely correlated with the energy demand, especially in the first two iterations. The convergence is significantly accelerated with the sensitivity considered after the third iteration. The iteration process reveals that high LMP caused by heavy load or congestion in TS can stimulate large amounts of DERs in DSs to increase generation to reduce the total demand and the operation cost of DSs. In turn, the decreased DS demand alleviates the heavy load and congestion. Accordingly, LMP and the operation cost in TS will fall. The DS demand in TS scheduling is equal to the actual demand of DSs after self-scheduling through the coordination. Power balance is satisfied at the PSP and energy mismatch is eliminated. Consequently, DERs in DSs are fully utilized to reduce the operation costs and electricity prices, as well as improve the system operation in TS and DSs. The balance of energy and interests is realized through TS-DS coordination.

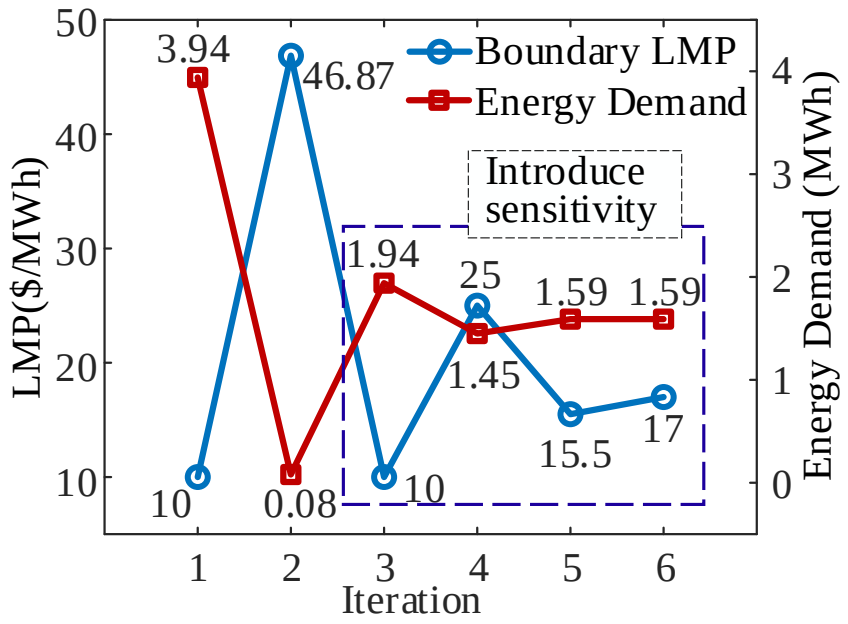

Fig.10. Convergency performance in a DS at bus D.

The parameters of the T118D69 system are given in [24]. It requires 10 iterations and 804.32 seconds. Thus, the proposed method meets calculation requirements in day-ahead market.

## VII. Conclusion

This paper proposes a novel day-ahead DEM clearing and pricing mechanism considering the uncertainty of RDG and the coordination with the WEM. The effectiveness of the proposed method has been verified by case studies and some conclusions can be drawn: 1) DLMPs internalize the uncertainty of RDG into the marginal costs of energy, voltage, congestion, and loss. Thus, the proposed DEM pricing method can provide effective price signals for managing not only the voltage, congestion, and loss, but also uncertainty. 2) High uncertainty of RDG reduces their profits, and increases electricity prices and operation costs in WEM and DEM. DLMPs can stimulate uncertainty sources to improve forecast accuracy. 3) The WEM-DEM coordination contributes to congestion mitigation and reduction of electricity prices. The total costs can be reduced by 3.38% in TS and 29.39% in DSs as compared with the separate case.

The proposed method provides a way to integrate WEM and DEM into a unified power market. LMP/ULMP and DLMPs are similar in form and can provide effective price signals for settlement in electricity markets considering uncertainties. This novel market mechanism establishes a platform for DERs to participate in the WEM based on the derived DLMPs.